\documentclass[conference]{IEEEtran}

\usepackage{graphicx}
\usepackage{graphics}
\usepackage{mathptmx} 
\usepackage{times} 
\usepackage{mathdots}
\usepackage{array}
\usepackage{amsmath, amsthm, amssymb, fancyhdr, lastpage}
\usepackage{nicefrac}
\usepackage{cite}

\ifCLASSINFOpdf
\else
\fi
\IEEEoverridecommandlockouts

\begin{document}

\title{Load Synchronization and Sustained Oscillations Induced by
  Transactive Control}

\author{\IEEEauthorblockN{Md Salman Nazir \hspace{2cm}Ian A.~Hiskens}
	\IEEEauthorblockA{Department of Electrical Engineering and
          Computer Science\\
		University of Michigan \\
		Email: mdsnazir,hiskens@umich.edu}

\thanks{The authors gratefully acknowledge the contributions of the
  Natural Sciences and Engineering Research Council of Canada (NSERC)
  and the U.S.~National Science Foundation through grant
  CNS-1238962.}}

\maketitle

\begin{abstract}
Transactive or market-based coordination strategies have recently been
proposed for controlling the aggregate demand of a large number of
electric loads. Such schemes offer operational benefits such as
enforcing distribution feeder capacity limits and providing users with
flexibility to consume energy based on the price they are willing to
pay. However, this paper demonstrates that they are also prone to load
synchronization and power oscillations. A transactive energy framework
has been adopted and applied to a population of thermostatically
controlled loads (TCLs). A modified TCL switching logic takes into
account market coordination signals, alongside the natural
hysteresis-based switching conditions. Studies of this market-based
coordination scheme suggest that several factors may contribute to
load synchronism, including sharp changes in the market prices that
are broadcast to loads, lack of diversity in user specified bid
curves, low feeder limits that are encountered periodically, and the
form of user bid curves. Case studies illustrate challenges associated
with market-based coordination strategies and provide insights into
modifications that address those issues.
\end{abstract}


%
\IEEEpeerreviewmaketitle

%

\section{Introduction}
	
The modeling and control of electric loads and their applications to
power systems services have been considered in various studies
\cite{Callaway2010}. Due to the significant potential of
thermostatically controlled loads (TCLs) (e.g.~air-conditioners, space
and water heaters), several control techniques have been explored in
the literature, with applications ranging from fast regulation or load
following \cite{Callaway2009,Bashash2013,Mathieu2013}, to optimizing
the day-ahead generation schedules \cite{Nazir2016}. These strategies
typically use either direct control via set-point variation
\cite{Callaway2009,Kundu2011} or probabilistic switching-based
distributed control \cite{Mathieu2013}. Another stream of recent work
employs market-based coordination strategies or a so-called
transactive energy control framework to manage the aggregate demand of
a large number of electric loads \cite{Jin2012, Huang2010, Hao2016,
  Li2016a,Li2016b, Bejestani2014, Knudsen2016}. Risks related to load
synchronization and cold load pick up are discussed in
\cite{Ihara1981, Kundu2012,Perfumo2013a} in the context of direct load
control, but there has been limited work to investigate such risks
under market coordination strategies. Hence, the objective of this
paper is to identify cases where oscillatory behavior of either power
consumption levels or clearing price may emerge when market-based
coordination signals are used, and to investigate the factors that
give rise to such behavior.

Transactive control (TC) demonstration projects have shown that with
residential loads market-based coordination strategies can reduce
utility demand and congestion at peak times \cite{Huang2010,
  Jin2012,Li2016a, Li2016b}. An optimization problem has been
formulated in \cite{Li2016a} where the coordinator first makes control
decisions to maximize the social welfare, and then the individual
users choose energy consumption to maximize individual utility based
on the coordinator's control decisions. The companion paper
\cite{Li2016b} demonstrates the applicability of the proposed
approach. However, the impact of control strategies on the temperature
dynamics, as well as the possibility and causes of power oscillations,
have not been analyzed in these studies. This paper
investigates factors that could lead to oscillatory response.

To achieve this objective, the transactive coordination mechanism
\cite{Jin2012,Li2016a,Li2016b} has been adopoted and applied to a
population of TCLs. We present a modified TCL switching logic that
takes into account market coordination signals, alongside the natural
switching conditions. Simulations suggest that several factors could
contribute to load synchronization and power oscillations, including
sharp changes in market prices broadcast to loads, lack of diversity
in user specified bid curves, feeder limits being set too low and
being encountered periodically, and the form of user bid curves. The
case studies illustrate challenges associated with market-based
coordination and control strategies.  The insights obtained through
these investigations provide a basis for addressing these challenges
through modifications to the control and market mechanisms.

	
\section{Modeling TCLs in a Transactive Control
  Framework}\label{sec:Model}

\subsection{TCL model preliminaries}

Consider a large population of TCLs. The set-point, dead-band,
internal and ambient temperatures ($^\circ$C) corresponding to each
load $i$ are denoted by $\theta^{\textrm{s}}_{i}$, $\delta_{i}$,
$\theta_i$ and $\theta^{\textrm{a}}$, respectively. Each load can be
modeled as a thermal capacitance, $C_i$ (kWh/$^\circ$C), in series
with a thermal resistance, $R_i$ ($^\circ$C/kW). Finally, the binary
variable $m_i$ denotes whether the load is on or off, and $P_i$ (kW)
the energy transfer rate when a cooling (or heating) TCL is switched
ON\@. One can model the dynamics of TCLs using a set of independent
first-order difference equations \cite{Mortensen1988},
\begin{equation} \label{eq:TCL_DE}
\theta_{i, t+h} = a_{i}\theta_{i,t}+(1-a_{i}) (\theta^{\textrm{a}}-m_{i,t}
\theta^{\textrm{g}}_{i}) + w_{i,t}
\end{equation}
where $h$ is the time-step, $a_i = e^{\nicefrac{-h}{(C_i R_i)}}$ is
the parameter governing the thermal characteristics of the thermal
mass, $\theta^{\textrm{g}}_{i}= P_i R_i$ is the temperature gain when
a cooling TCL is ON and $w$ is a noise process. The variable $m_{i,t}$
for TCL $i$ captures the TCL's switching behavior according to,
\begin{equation}
m_{i,t+h}=
\begin{cases}
0, & \text{if} \quad \theta_i < \theta^{\textrm{min}}_{i}\\
1, & \text{if} \quad \theta_i > \theta^{\textrm{max}}_{i}\\
m_{i,t}, & \text{otherwise}
\end{cases}
\end{equation}
where $\theta^{\textrm{min}}_{i} = \theta^{\textrm{s}}_{i} - \delta_i/2$ and $\theta^{\textrm{max}}_{i} = \theta^{\textrm{s}}_{i} + \delta_i/2$.

With coefficient of performance (scaling factor related to efficiency
\cite{Callaway2009}), $\eta_i$, the aggregate electrical power
consumed by $N_{\textrm{TCL}}$ devices is given by,
\begin{equation}
P^{\textrm{elec}}_{t} = \sum_{i=1}^{N_{\textrm{TCL}}}{\nicefrac{m_{i,t} P_i}{ \eta_i}}.
\end{equation}

\subsection{Transactive coordination framework}


The transactive control framework is based on a double auction
mechanism \cite{Fuller2011}. Following the existing literature on the
TC framework and modeling of the market clearing mechanism
\cite{Li2016a}, \cite{Li2016b}, subsequent work is based on the
following assumptions: (i) A `coordinator' is present to receive the
bidding information from a population of devices and to send back the
market clearing information. (ii) Each device is equipped with a smart
thermostat that can measure the room temperature. It also has
communication capabilities to exchange bid information with the
coordinator. (iii) Before each market period, the device measures its
room temperature, and submits a bid to the coordinator. The bid should
consist of the load power and the bidding price. (iv) The device has
prediction capability to forecast its temperature 5~minutes ahead,
which it then uses to establish its bidding price. Hence, the bidding
price depends on the current temperature and the temperature 5~minutes
ahead.
	
In a TC framework, every load submits a demand bid where it specifies
its desired amount of energy demand over a specific interval. Note
that to be consistent with the literature, market clearing intervals
with \textit{5-minute duration} have been considered. Hence, the bids
are also based on average energy demand over 5-minute intervals.

\subsection{Modeling TCL bids} 

Based on the above framework, let $p^{\textrm{bid}}_{i,t}$ denote the
price bid of load $i$ at time $t$ and $q^{\textrm{bid}}_{i,t}$ be its
corresponding amount of energy demand over the next 5-minute period.

\begin{figure}[t]
\begin{center}
\includegraphics[scale=0.4]{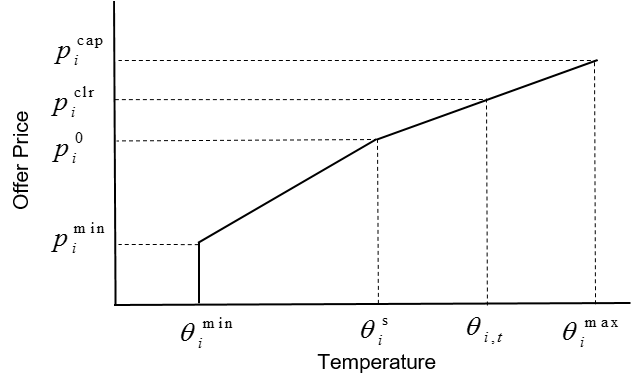}
\vspace{-2mm}
\caption{Demand side offer mapped to temperature.} \label{fig:offerVtemp2}
\end{center}
\end{figure}

Fig.~\ref{fig:offerVtemp2} shows how a TCL determines its bid
\cite{Li2016b,Fuller2011}. Here, an air-conditioner user bids $p^0_i$
if its temperature $\theta_{i,t}$ is at its set-point,
$\theta_i^{\textrm{s}}$ (i.e.~desired temperature level), with the
offer varying if the temperature deviates from
$\theta_i^{\textrm{s}}$. Above a certain threshold
$\theta_{i}^{\textrm{max}}$ the maximum bid is capped at
$p^{\textrm{cap}}_i$. Similarly, below the threshold
$\theta_{i}^{\textrm{min}}$ the TCL might not be willing to bid, so
places $p^{\textrm{bid}}_i = 0$. Fig.~\ref{fig:offerVtemp2} shows a
piecewise linear mapping, with slopes $\gamma_1$ and $\gamma_2$
depending on if the temperature is above or below the set-point.
Thus, the bid and temperature relation can be expressed as,
\begin{equation}
p^{\textrm{bid}}_{i,t}=
\begin{cases}
	(\theta_{i,t} - \theta^{\textrm{s}}_{i})\gamma_1 + p^0_i, & \text{if} \quad  \theta_{i,t} >=  \theta^{\textrm{s}}_{i} \\
	(\theta^{\textrm{s}}_{i}-\theta_{i,t})\gamma_2 + p^0_i, & \text{if} \quad  \theta_{i,t} <  \theta^{\textrm{s}}_{i} \\
	0, & \theta_{i,t} <  \theta^{\textrm{min}} \\
	p^{\textrm{cap}}_i, & \theta_{i,t} >  \theta^{\textrm{max}}.
\end{cases}
\end{equation}
Since the bids are over 5-minute intervals, whereas TCLs have faster
dynamics (few seconds), $\theta_{i,t}$ may be the latest measured
temperature, or a predicted temperature (e.g. at 2.5~minutes ahead)
based on its current on/off operating state, as detailed in
\cite{Li2016b}. Finally, $q^{\textrm{bid}}_{i,t}$ will be the average
power consumed if TCL $i$ remains on during the 5-minute interval.

\subsection{Market clearing mechanism}\label{sec:marketclear}

Since in a transactive framework the grid is organized in a
hierarchical way, the TCL nodes are connected to a distribution
feeder, which clears an allowable demand level at a particular
price. Initially the feeder broadcasts a base price, but adjusts that
price if the feeder capacity constraint is exceeded.

\begin{figure}[t]
\begin{center}
\includegraphics[scale=0.27]{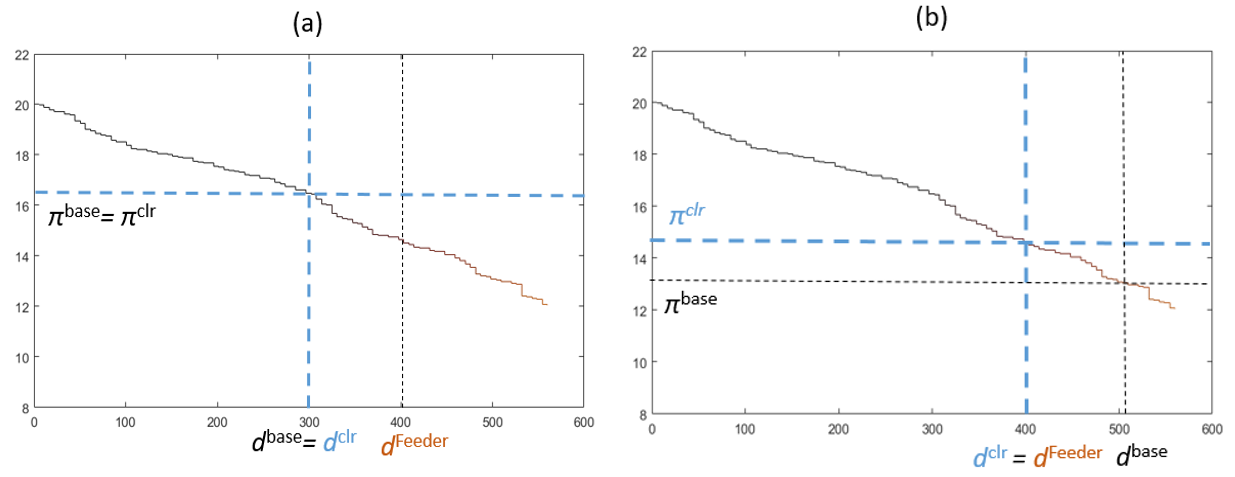}
\caption{(a) Market clearing with feeder capacity not exceeded. (b)
  Market clearing with feeder capacity exceeded.} \label{fig:TCclearing1}
\end{center}
\end{figure}

Let $\pi^{\textrm{base}}_{t}$ be the base price forecast at time $t$
and $d^{\textrm{base}}_t$ be the corresponding base aggregate
demand. The clearing price $\pi^{\textrm{clr}}_t$ and the cleared
aggregate demand $d^{\textrm{clr}}_{t}$ can be found at time $t$
according to the following algorithm, keeping in mind that
$d^{\textrm{clr}}_{t}$ must satisfy the feeder capacity limit,
\begin{equation}{\label{eq:feederCap}}
	d^{\textrm{clr}}_{t} \le d^{\textrm{Feeder}}. 
\end{equation}
The overall transactive control mechanism, based on
\cite{Li2016a,Li2016b}, can be summarized as:
\begin{enumerate}
\item Gather anonymous bids (price versus demand) and build an aggregate
  demand function (see Fig.~\ref{fig:TCclearing1}).
  
\item Using the aggregate demand function and the base price
  information for that time period $\pi^{\textrm{base}}_t$, obtain the
  corresponding base aggregate demand $d^{\textrm{base}}_t$.

\item If $d^{\textrm{base}}_t < d^{\textrm{Feeder}}$ (see
  Fig.~\ref{fig:TCclearing1}(a)), $d^{\textrm{clr}}_{t}$ =
  $d^{\textrm{Feeder}}$. Set $\pi^{\textrm{clr}}_t$ =
  $\pi^{\textrm{base}}_{t}$.

\item If $d^{\textrm{base}}_t \ge= d^{\textrm{Feeder}}$ (see
  Fig.~\ref{fig:TCclearing1}(b)), set $d^{\textrm{clr}}_{t}$ =
  $d^{\textrm{Feeder}}$. Set $\pi^{\textrm{clr}}_t$.

\item Each load compares its offer with $\pi^{\textrm{clr}}_t$ and
  self-dispatches if $\pi^{\textrm{bid}}_{i,t} \ge
  \pi^{\textrm{clr}}_t$.
\end{enumerate}
Note that the above market clearing mechanism ignores the network
structure and the network flow constraints \cite{Li2016a}.

\subsection{Modified TCL switching logic}

Under the transactive framework, the switching variable $m_{i,t}$ in
(\ref{eq:TCL_DE}) will be multiplied by an additional decision
variable $v_{i,t}$, thus the overall expression becomes,
\begin{equation} \label{eq:TCL_DE_mod}
\theta_{i,t+h} = a_{i} \; \theta_{i,t}+(1-a_{i})
(\theta^{\textrm{a}}-m_{i,t} \; v_{i,t} \; \theta^{\textrm{g}}_i)
\end{equation} 
where,  
\begin{equation}
v_{i,t}=
\begin{cases}
0, & \text{if} \quad  \pi^{\textrm{bid}}_{i,t}  < \pi^{\textrm{clr}}_t \\
1, & \text{if} \quad  \pi^{\textrm{bid}}_{i,t} \ge \pi^{\textrm{clr}}_t. \\
\end{cases}
\end{equation}
Here, $v_{i,t}$ can be thought of as an upper level decision variable,
the TCL's response to a transactive incentive signal or a clearing
price $\pi^{\textrm{clr}}_t$. If at any time $v_{i,t} = 1$ then the
TCL simply follows its natural thermostat cycle. Note that under the
above switching scheme,
\begin{enumerate}
\item A TCL consumes power when $m_{i,t} = 1$, and $v_{i,t} = 1$.
\item A TCL does not consume power when $m_{i,t} = 1$, $v_{i,t} = 0$. 
\item A TCL does not consume power when $m_{i,t} = 0$ (natural
  thermostat off mode).
\end{enumerate}

\section{Case Study}
	
Consider a population of 1000 TCLs. Parameter values are similar to
those used in \cite{Callaway2009 ,Bashash2013}. A base price is sent
at 5-minute intervals.  The coordinator sends the participants only
the 5-minute ahead base price. Each load's bid levels are constructed
with continuous offers, similar to Fig.~\ref{fig:offerVtemp2}. Bid
levels can range between 10 to 50~\$/MWh. Each load has its own slopes
$\gamma_1$ and $\gamma_2$ for its bid curve. Additionally, the feeder
capacity constraint was set at 70$\%$ of the maximum power capacity of
the TCLs (5600~kW for 1000 TCLs). Since the simulation of TCL
temperature dynamics requires faster time steps, while market clearing
occurs every 5~minutes, the TCL temperature dynamics were simulated
using a time-step of $h=10$~s, and the market mechanisms were
simulated with 5-minute time-steps.

\subsection{Oscillations induced due to changes in base price}

Initial investigations considered the response of TCLs to sharp
changes in the base price. The base price is initially 42~\$/MWh and
stays at that level for 6 hours before suddenly dropping to 20~\$/MWh
for a further 6~hours, and then finally to 9~\$/MWh for the remainder
of the time. In reality, these price changes might correspond to
sudden changes in background demand, such as an industrial load or
electric vehicle charging.

\begin{figure}[t]
\begin{center}
\includegraphics[scale=0.43]{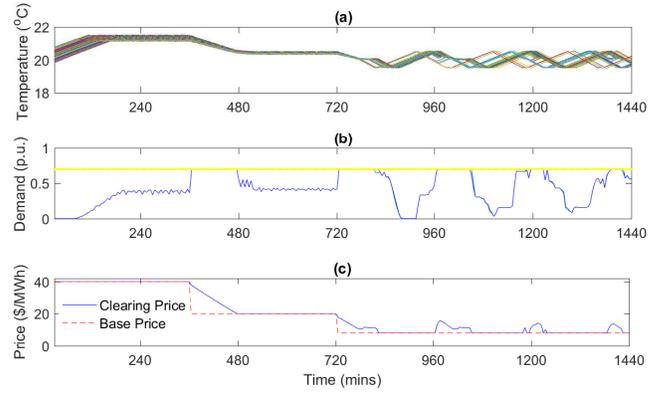}
\caption{(a) Temperature evolution of individual TCLs, (b) 5-minute
  average aggregate demand, (c) base price and clearing price.}
			\label{fig:ncase3}
\end{center}
\end{figure}

Fig.~\ref{fig:ncase3} provides a prototypical example of TCL
synchronization. The TCLs started with diverse initial
temperatures. But because the base price remained high (at 42~\$/MWh)
for a few hours, most of the TCLs did not initially consume
power. (Their bids were not sufficiently high to be cleared.) However,
within a few hours (around minute 200) their temperatures
synchronized. Later, as the base price drops to 20~\$/MWh, TCLs find
the price level favorable and want to turn on. The aggregate demand
reaches the feeder limit, causing the market clearing price to rise
above the base price level. During minutes 480-720, the demand stays
flat and TCL temperatures remain close to their set-point
values. Next, at $t$=720~min, when the price drops to 9~\$/MWh, the
TCLs find this low price even more favorable and many compete to
consume power. Large oscillations in aggregate power are observed and
the feeder limit is hit periodically. Thus, a step change in base
price, especially to a low value, can induce large power
oscillations. This is mainly due to TCL temperatures becoming
synchronized during preceding periods of relatively high base prices.

\begin{figure}[t]
\begin{center}
\includegraphics[scale=0.30]{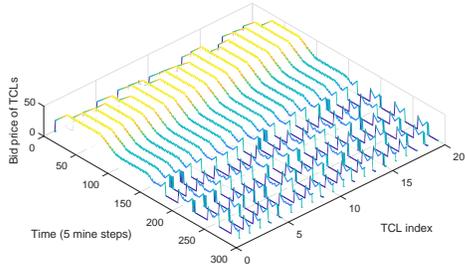}
\caption{Bid evolution at 5 minute intervals for 20 randomly chosen TCLs.} 
			\label{fig:ncase3waterfall1}
\end{center}
\end{figure}

Fig.~\ref{fig:ncase3waterfall1} shows the evolution of bids for 20
TCLs (with 5~minute time-steps on the x-axis).  Once synchronized,
groups of similar bids are cleared and so those TCLs begin to cool. As
they cool, their bids fall, allowing other groups with higher bids to
be cleared.

\begin{figure}[t]
\begin{center}
\includegraphics[scale=0.428]{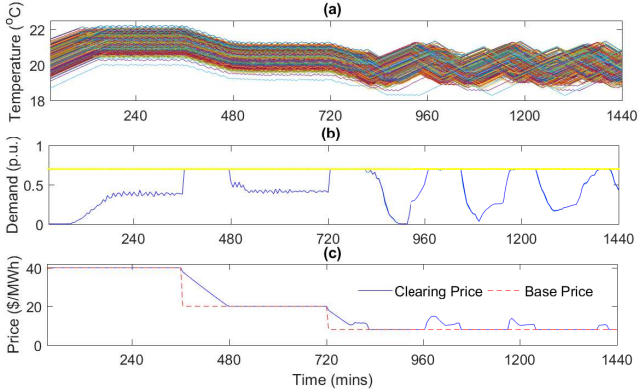}
\caption{Oscillations in 5-minute average aggregate demand, induced by
  sharp changes in base price, despite heterogeneity in TCL
  set-points.} \label{fig:case2a}
\end{center}	
\end{figure}
		
Besides heterogeneity in bid curves, customers may also have different
set-points for their individual air-conditioners. While studies show
that heterogeneity in the population leads to damping of oscillations
under direct load control \cite{Perfumo2013a}, Fig.~\ref{fig:case2a}
shows that step changes in the base price still result in large
oscillations. Results are similar to the case without heterogeneity in
the set-points (Fig.~\ref{fig:ncase3}). This is understandable because
even though set-points vary, the relative temperature differences
(compared to the individual set-points) may still synchronize, which
then leads to oscillations in aggregate demand when the base price
falls considerably.

\subsection{Fast transients due to temperature synchronization and
  fluctuating prices}

\begin{figure}[t]
\begin{center}
\includegraphics[scale=0.43]{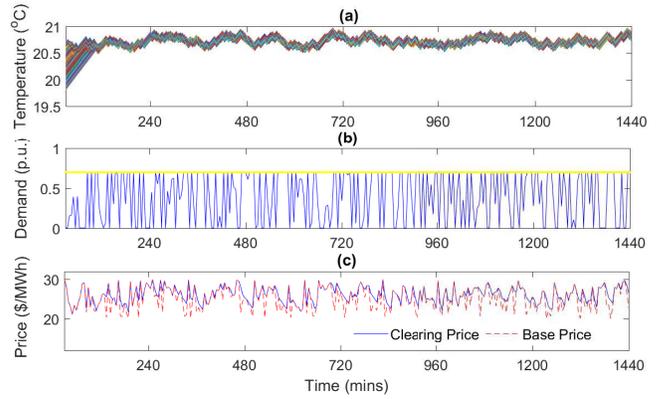}
\caption{Temperature synchronization and highly fluctuating demand due
  to moderate fluctuations in base price.} \label{fig:ncase2}
\end{center}
\end{figure}

Instead of large step changes in price, this case considered a price
signal which fluctuates between 20 and 30~\$/MWh. Behavior is shown in
Fig.~\ref{fig:ncase2}. Surprisingly this triggers a highly fluctuating
response in the 5-minute average TCL demand. Investigations suggest that
variations in the TCL bids (as their temperatures change) relative to
the base price cause these sharp transients in aggregate power levels.

This study assumed that the slopes of the bid curves, though
heterogeneous, are not significantly different. Initially very few TCL
bids were sufficiently high to be cleared. Hence, their temperatures
rose to around $20.6^\circ$C\@. At this point, many placed
sufficiently high bids and were subsequently cleared. If the base
price remained unchanged, these TCLs would continue to consume power
enabling their temperatures to reach the desired set-points. However,
if the base price were to rise slightly, it would cause some TCLs to
turn off since their bids become unfavorable. Conversely, if the
majority of the TCLs were off, then a small drop in the base price
would lead to TCLs with similar bids being cleared and turning on. As
their temperatures approach their set-points, they bid lower and at some
point will no longer be cleared. Thus, these relative movements of the
TCL bids (due to changes in their temperatures) compared to the base
price levels may lead to significant fluctuations in the aggregate
power, as shown in Fig.~\ref{fig:ncase2}(b).

\subsection{Oscillations induced due to feeder capacity constraint}

\begin{figure}[t]
\begin{center}
\includegraphics[scale=0.41]{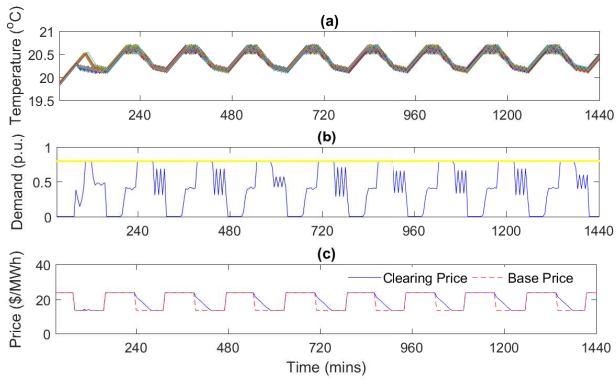}
\caption{Fast oscillations due to groups of TCLs having synchronized
  temperatures.} \label{fig:ncase4}
\end{center}
\end{figure}

Fig.~\ref{fig:ncase4} shows a situation where fast oscillations were
induced due to the feeder capacity constraint. The base price signal
in this case resembles a pulse train fluctuating between 14 and
24~\$/MWh. Every time the base price drops, TCLs switch on and the
base aggregate demand of the TCLs reaches the feeder limit. For
example, when the price drops to 14~\$/MWh at $t$ = 240~min, all TCLs
want to cool since their temperatures have risen considerably during
the preceding high price period. However, if all TCLs turn on at the
same time, the feeder limit will be violated. Following the mechanism
described in Section \ref{sec:marketclear}, the clearing price is
revised above the base price and therefore feeder limits are
respected. However, as the clearing price approaches 14~\$/MWh, a
specific pattern of fast oscillations emerges, as seen in
Fig.~\ref{fig:ncase4}(b).

By the time clearing prices approach 14~\$/MWh, TCL temperatures are
near their set-points so they offer low bids. However, a fraction of
TCLs still bid higher than 14~\$/MWh and are cleared. As these cool
more, they bid lower and subsequently turn off. By that time,
temperatures of a second group have risen such that their bids now
exceed 14~\$/MWh and they turn on. Thus, the most aggressive ones get
cleared first, then the next group, and so on. Subsequently, as the
base price rises again to 24~\$/MWh, all loads turn off since they are
unwilling to pay such a high price when their temperatures are already
near their desired set-points. This behavior continues as long as the
base price keeps oscillating.
 
\subsection{Oscillations due to subgroups of TCLs with similar bid curves}	

\begin{figure}[t]
\begin{center}
\includegraphics[scale=0.43]{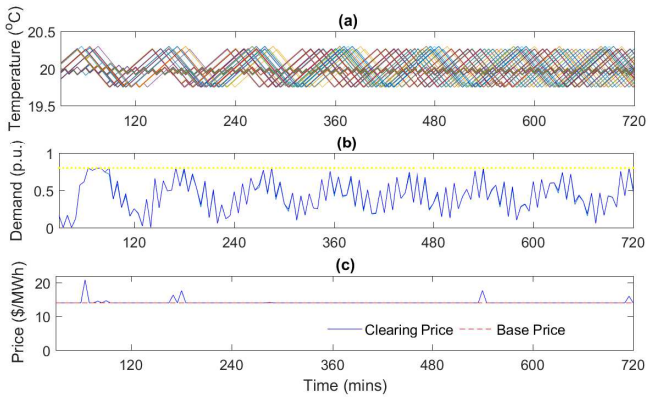}
\caption{Fast power oscillations due to groups of TCLs having
  similar bids, leading to their synchronized temperatures.} \label{fig:Case6}
\end{center}
\end{figure}

This case shows that it is not necessary for all TCLs to be
synchronized at the same temperature to cause power oscillations. It
can be seen from Fig.~\ref{fig:Case6} that groups of TCLs have
synchronized temperatures, with TCLs within each group evolving in a
similar manner. This then results in quasi-periodic behavior for the
ensemble of loads. Besides large magnitude oscillations in power, the
ensemble demand also displays jitter. The quasi-periodic evolution of
the ensemble results in mixing of oscillations of different
frequencies.

\section{Discussions and Future Work}

A transactive coordination mechanism has been applied to a population
of TCLs. A modification to TCL switching logic was established to take
into account market coordination signals, alongside the natural
hysteresis-based switching of TCLs. Investigations identified
conditions that give rise to load synchronization and power
oscillations. Simulations suggest that several factors can contribute
to such synchronism, including sharp changes in base price, prolonged
flat base prices, lack of diversity in user specified bid curves, the
form of the bid curves, and similarity of bid curves across subgroups
of TCLs. It was also observed that imposing a feeder limit constraint,
while effectively limiting demand through adjustment of market
clearing prices, may lead to an oscillatory power response where
jitters appear due to mixing of different frequency oscillations from
groups of separately synchronized TCLs. Future research will
investigate these effects in a more formal Poincar\'e analysis setting
and develop control algorithms that are able to avoid the risks of
oscillatory behavior from synchronized TCLs.

 \bibliographystyle{IEEEtran}
 \bibliography{PEStransactive}

\begin{thebibliography}{10}
\providecommand{\url}[1]{#1}
\csname url@samestyle\endcsname
\providecommand{\newblock}{\relax}
\providecommand{\bibinfo}[2]{#2}
\providecommand{\BIBentrySTDinterwordspacing}{\spaceskip=0pt\relax}
\providecommand{\BIBentryALTinterwordstretchfactor}{4}
\providecommand{\BIBentryALTinterwordspacing}{\spaceskip=\fontdimen2\font plus
\BIBentryALTinterwordstretchfactor\fontdimen3\font minus
  \fontdimen4\font\relax}
\providecommand{\BIBforeignlanguage}[2]{{%
\expandafter\ifx\csname l@#1\endcsname\relax
\typeout{** WARNING: IEEEtran.bst: No hyphenation pattern has been}%
\typeout{** loaded for the language `#1'. Using the pattern for}%
\typeout{** the default language instead.}%
\else
\language=\csname l@#1\endcsname
\fi
#2}}
\providecommand{\BIBdecl}{\relax}
\BIBdecl

\bibitem{Callaway2010}
D.~S. Callaway and I.~A. Hiskens, ``Achieving controllability of electric
  loads,'' \emph{Proceedings of the IEEE}, vol.~99, no.~1, pp. 184--199, Jan
  2011.

\bibitem{Callaway2009}
D.~S. Callaway, ``{Tapping the energy storage potential in electric loads to
  deliver load following and regulation, with application to wind energy},''
  \emph{Energy Conversion and Management}, vol.~50, no.~5, pp. 1389--1400,
  2009.

\bibitem{Bashash2013}
S.~Bashash and H.~K. Fathy, ``{Modeling and control of aggregate air
  conditioning loads for robust renewable power management},'' \emph{IEEE
  Transactions on Control Systems Technology}, vol.~21, no.~4, pp. 1318--1327,
  2013.

\bibitem{Mathieu2013}
J.~L. Mathieu, S.~Koch, and D.~S. Callaway, ``{State estimation and control of
  electric loads to manage real-time energy imbalance},'' \emph{IEEE
  Transactions on Power Systems}, vol.~28, no.~1, pp. 430--440, 2013.

\bibitem{Nazir2016}
M.~S. Nazir, F.~D. Galiana, and A.~Prieur, ``{Unit Commitment Incorporating
  Histogram Control of Electric Loads with Energy Storage},'' \emph{IEEE
  Transactions on Power Systems}, vol.~31, no.~4, pp. 2857--2866, 2016.

\bibitem{Kundu2011}
S.~Kundu, N.~Sinitsyn, S.~Backhaus, and I.~Hiskens, ``{Modeling and Control of
  Thermostatically Controlled Loads},'' in \emph{Proceedings of the 17th Power
  Systems Computation Conference}, 2011.

\bibitem{Jin2012}
D.~Jin, X.~Zhang, and S.~Ghosh, ``{Simulation models for evaluation of network
  design and hierarchical transactive control mechanisms in smart grids},'' in
  \emph{2012 IEEE PES Innovative Smart Grid Technologies}, 2012, pp. 1--8.

\bibitem{Huang2010}
P.~Huang, J.~Kalagnanam, R.~Natarajan, M.~Sharma, R.~Ambrosio, D.~Hammerstrom,
  and R.~Melton, ``{Analytics and Transactive Control Design for the Pacific
  Northwest Smart Grid Demonstration Project},'' in \emph{Smart Grid
  Communications (SmartGridComm), 2010 First IEEE International Conference on},
  2010, pp. 449--454.

\bibitem{Hao2016}
H.~Hao, C.~D. Corbin, K.~Kalsi, and R.~G. Pratt, ``{Transactive Control of
  Commercial Buildings for Demand Response},'' \emph{IEEE Transactions on Power
  Systems}, pp. 1--1, 2016.

\bibitem{Li2016a}
S.~Li, W.~Zhang, J.~Lian, and K.~Kalsi, ``{Market-Based Coordination of
  Thermostatically Controlled Loads - Part I: A Mechanism Design
  Formulation},'' \emph{IEEE Transactions on Power Systems}, vol.~31, no.~2,
  pp. 1170--1178, 2016.

\bibitem{Li2016b}
------, ``{Market-Based Coordination of Thermostatically Controlled Loads -
  Part II: Unknown Parameters and Case Studies},'' \emph{IEEE Transactions on
  Power Systems}, vol.~31, no.~2, pp. 1--9, 2016.

\bibitem{Bejestani2014}
A.~K. Bejestani, A.~Annaswamy, and T.~Samad, ``{A hierarchical transactive
  control architecture for renewables integration in smart grids: Analytical
  modeling and stability},'' \emph{IEEE Transactions on Smart Grid}, vol.~5,
  no.~4, pp. 2054--2065, 2014.

\bibitem{Knudsen2016}
J.~Knudsen, J.~Hansen, and A.~M. Annaswamy, ``{A Dynamic Market Mechanism for
  the Integration of Renewables and Demand Response},'' \emph{IEEE Transactions
  on Control Systems Technology}, vol.~24, no.~3, pp. 940--955, 2016.

\bibitem{Ihara1981}
S.~Ihara and F.~C. Schweppe, ``{Physically based modeling of cold load
  pickup},'' \emph{IEEE Transactions on Power Apparatus and Systems}, vol.
  PAS-100, no.~9, pp. 4142--4150, 1981.

\bibitem{Kundu2012}
S.~Kundu and N.~Sinitsyn, ``Safe protocol for controlling power consumption by
  a heterogeneous population of loads,'' in \emph{2012 American Control
  Conference (ACC)}, June 2012, pp. 2947--2952.

\bibitem{Perfumo2013a}
C.~Perfumo, J.~Braslavsky, J.~K. Ward, and E.~Kofman, ``{An analytical
  characterisation of cold-load pickup oscillations in thermostatically
  controlled loads},'' in \emph{2013 3rd Australian Control Conference}, 2013,
  pp. 195--200.

\bibitem{Mortensen1988}
R.~E. Mortensen and K.~P. Haggerty, ``{Stochastic computer model for heating
  and cooling loads},'' \emph{IEEE Transactions on Power Systems}, vol.~3,
  no.~3, pp. 1213--1219, 1988.

\bibitem{Fuller2011}
J.~C. Fuller, K.~P. Schneider, and D.~Chassin, ``{Analysis of residential
  demand response and double-auction markets},'' in \emph{IEEE Power and Energy
  Society General Meeting}, 2011, pp. 1--7.

\end{thebibliography}
\end{document}